\documentclass[a4paper]{JAC2003}
\usepackage{graphicx}
\usepackage{booktabs}
\usepackage{xcolor}
\usepackage{url}
\usepackage{microtype}

\usepackage{varwidth}

\usepackage{xcolor}

\newcommand\query[1]{}%

\begin{document}
\title{BTF MEASUREMENTS WITH BEAM--BEAM INTERACTIONS}

\author{P. G\"orgen, O. Boine-Frankenheim, TEMF, TU-Darmstadt, Germany\\ W. Fischer, Brookhaven National Laboratory, Upton, NY, USA}

\maketitle

\begin{abstract}
We present considerations about the transverse beam transfer function (BTF) of beams under the influence of two effects: The strong--strong beam--beam effect and the influence of a Gaussian electron lens. The BTF are investigated using two methods: BTF excitation is simulated in a particle-in-cell (PIC) code. The BTF model is verified using a known analytic expectation.  Analytic expectations for BTF of beams under a stationary electron lens are derived by extending BTF from the formalism of Berg and Ruggiero. Finally we compare the analytic BTF results for a stationary Gaussian lens to both the PIC simulation for split tune conditions and to PIC simulations for a beam influenced by an electron lens. We conclude that the formalism represents the electron lens well and can be applied to a limited extend to the beam--beam effect under split tune conditions. The analytic formalism allows us to recover the strength of an electron lens by means of fitting and can give clues regarding the strength of the beam--beam effect under split tune conditions.
\end{abstract}

\section{Reconstruction of tune spread from BTF}
For a long time there has been a desire at BNL to recover the beam--beam parameter and with it the tune spreads from BTF of beams undergoing the beam--beam effect. Recently this desire has been intensified by the construction and installation of the electron lens~\cite{ElensProgress2012}. Normally the machine is run in conditions with the two rings tuned to identical or near-identical tunes. We refer to these conditions as \emph{normal conditions}. Under these circumstances, the coherent beam--beam modes often dominate the BTF. When observed, the distance between $\pi$ and $\sigma$ modes can be used to determine the beam--beam parameter and with it one can estimate the tune spread. For diagnosing the tune spread due to the electron lens we can not expect $\pi$ and $\sigma$  modes outside the incoherent spectrum: The electron beam is dumped after usage and not fed back into the system. However, running the electron lens will lead to a tune spread similar to the one caused by the beam--beam effect (but with a positive tune shift). This tune spread in turn will lead to a deformation of the betatron peak. We would like to be able to recover the strength of the electron lens from this deformation in the absence of beam--beam interactions.

A similar situation can be hoped for in runs with split tunes. We talk about split tunes when the tunes are, unlike normal conditions, offset in the two rings; for example, during the 2012 split tunes run they were typically separated by about 0.04 and located on either side of the 7/10 resonance line. In this case the coherent lines can move into the incoherent spectrum where they can be landau-damped. The resulting beam heating has been observed in measurement~\cite{simonMeasure}. In simulation, the incoherent spectrum of the beam--beam effect of split tunes leads to BTF similar to those of an electron lens. However, due to the opposite sign of the force of the beam--beam and the electron lens the betatron peak is located on the other side of the lattice tune.

When talking about BTF we should specify what we mean exactly. The BTF system at RHIC uses the direct diode detection technique~\cite{bbq} developed at CERN together with a coherent excitation signal fed onto the beams. The complex response amplitude as a fraction of excitation amplitude gives the BTF at the frequency of the excitation. The excitation signal is swept over a range of frequencies around the fractional tune to obtain the BTF as a
function of frequency. Commonly the complex BTF is separated into phase and amplitude and the result is shown as a function of frequency.

\section{BTF of coasting beams}
The BTF of coasting beams have been studied in detail in the past, as they give, for example, information on the stability limits for beams under space charge. Examples of such studies can be found in \cite{burov09,kornilov07}. We think that the machine conditions in RHIC during pp-operation allow us to apply coasting beam considerations for the bunched beams in RHIC: the synchrotron period with the 28 MHz radio frequency (RF) system and the time used to take one frequency sample in the baseband-Q (BBQ) BTF system are about the same ($\sim$30~ms). For this reason, from the point of view of the BTF measurement, the beam might be expected to resemble a coasting beam. Longitudinal motion is very slow on the timescale of the BTF measurement. Simulations include chromaticity and synchrotron motion. It will be shown later that they agree with a coasting-beam description.

\subsection{Application to the BTF of Beams Under Beam--Beam or an Electron Lens}
To account for an electron lens or a beam--beam effect where the coherent modes lie inside the incoherent spectrum, we use the theory developed by Berg and Ruggiero in~\cite{berg}. We only have to extend the BTF they give by replacing the tune change due to an octupole with the tune change due to a Gaussian lens. The BTF by Berg and Ruggiero is
\begin{equation}
R(\Omega)=\int_0^\infty\int_0^\infty
\frac{1}{\Omega-\omega_x\left(J_x,J_y\right)}
\frac{J_x\,\mathrm d \psi_x}{\mathrm d J_x}
\psi_y\,\mathrm dJ_x \,\mathrm dJ_y,\label{eq:anaBTF}
\end{equation}
where $J_x$ and $J_y$ are the transverse action angle variables, $\psi_x,\psi_y$ are the distribution functions in action angle variables, separated into the contribution of $x$ and $y$ direction, $\omega_x(J_x,J_y)$ is betatron frequency as a function of these variables, and $\Omega$ is the frequency at which the BTF is calculated.

For the distribution functions $\psi_x,\psi_y$, we use the distribution of a Gaussian transverse distribution in action angle variables as found for example in \cite{ng} (we can also simply split the $\psi_0$ we find
in~\cite{berg} to get $\psi_x,\psi_y$). We also need the formula for the single particle tune shift due to a Gaussian lens $\Delta Q_\mathrm{lens}$ which as a function of the action-angle amplitudes of the particles $J_{x,y}$ and the peak tune shift $\Delta Q_0$ is~\cite{burov09}
\begin{equation}
\Delta Q_\mathrm{lens}=\Delta Q_0 \int_0^1\frac{\left(I_0
\left(\frac{J_xz}{2}\right)-I_1\left(\frac{J_x z}{2}\right)
\right)I_0\left(\frac{J_yz}{2}\right)}{\exp(z(J_x+J_y)/2)}\,\mathrm d z.
\label{eq:fullDQ}
\end{equation}
Because the Bessel functions are slow to evaluate when one tries to calculate eq.~(\ref{eq:anaBTF}) numerically, instead of this analytic expression we use a well-behaved replacement that is friendlier for numerics and was developed for the treatment of space charge in~\cite{burov09}. With $a_{x,y}=\sqrt{2 J_{x,y}}$, the approximation for our $\Delta Q_\mathrm{lens}$ is now
\begin{equation}
\Delta Q_0\frac{192-11a_x-18\sqrt{a_xa_y}+3a_y^2}
{192-11a_x-18\sqrt{a_xa_y}+3a_y^2+36a_x^2+24a_y^2} .\label{eq:simpleDQ}
\end{equation}
For the longitudinal distribution we assume a Gaussian momentum spread with the tune shift according to chromaticity resulting in a tune deviation $\Delta Q_\mathrm{chrom}$. It can be taken into account by modifying $R(\Omega)$ to include also the tune shift due to chromaticity, the resulting $R(\Omega)$ is
\begin{equation}
\int_{-\infty}^\infty\int_0^\infty\int_0^\infty
\frac{1}{\Omega-\omega_x\left(J_x,J_y,p\right)}
\frac{J_x\,\mathrm d \psi_x}{\mathrm d J_x}
\psi_{yp}\,\mathrm dJ_x \,\mathrm dJ_y\,\mathrm d p,\label{eq:fullanaBTF}
\end{equation}
where $\psi_{yp}=\psi_y(y)\psi_p(p)$ contains the combined densities in the vertical and momentum plane.
The resulting $\omega_x$ is`
\begin{equation}
\omega_x(J_x,J_y,p)=\omega_0\left(Q_0+\Delta Q_\mathrm{lens}(J_x,J_y)+\Delta Q_\mathrm{chrom}(p)\right)
\end{equation}
with $\omega_0$ the revolution frequency, $Q_0$ the lattice tune and $\Delta Q_\mathrm{chrom}(p)$ the tune shift due to chromaticity. The chromaticity usually plays a minor role for the BTF of realistic beams because RHIC runs at low chromaticity.
\section{Simulation model}
For the investigation of BTF of beams undergoing the beam--beam effect in RHIC, a simulation model was implemented on top of the particle tracking code \textsc{Patric}~\cite{patric}. For the tracking between interaction points~(IPs), matrices from madx are used. The translation between IPs is done by one single matrix multiplication with the linear one-turn map computed using the one-turn map on the 2012 100~GeV polarized proton lattice~\cite{lattice}. For synchrotron motion the respective parts of the madx result are ignored and replaced by a more versatile model which is present in the code; this allows us to take into account different RF waveforms. One instance of the code is run for each of the typically six (in the case of two IPs) or two (in the case of one IP) coupling bunches. The beam--beam interaction is taken into account by exchanging the two-dimensional electric fields between the bunches at the interaction points and kicking the particles accordingly. The fields are calculated using a two-dimensional fast-Fourier-transform-based Poisson solver with open boundary conditions~\cite{kapin}. The beam--beam implementation reproduces the expected behaviour, especially the $\pi$ an $\sigma$ modes are found at the expected positions of $Q$ and $Q-\lambda_\mathrm{yokoya}\cdot\xi_\mathrm{bb}$ with $Q$ the tune, $\xi_\mathrm{bb}$ the beam--beam parameter and $\lambda_\mathrm{yokoya}$ the Yokoya factor~\cite{yokoya}. The maximum single particle tune shift in simulation equals the beam--beam parameter as expected.

\subsection{BTF Implementation}
The BTF is implemented as follows. A particle ensemble of typically between $10^5$ and $10^7$ macroparticles is initialized as a matched Gaussian distribution and left coasting for a few thousand turns to equilibrate possible matching errors. After this initial equilibration the equilibrium distribution is cached. Then, a coherent excitation is carried out by adding a sinusoidal excitation signal $a(t)=\sin(\omega t)$ to the momentum component of the particle vectors at each passing of the exciter. Because the excitation frequency is chosen around the fractional tune, it is assumed that a whole bunch sees the same excitation signal. After each turn the transverse position of the centre of charge of the beam and the excitation signal amplitude is stored. In post-processing the BTF is calculated as the fraction of the complex amplitudes of the response and the exciting signal. The amplitudes are determined using the discrete Fourier transformation (DFT) at the chosen excitation frequency. After each excitation frequency the PIC-code re-initializes with the equilibrium particle distribution to reduce transient modes. To make sure we look primarily at the steady state of the excited beam, not at the transients, the first few hundred turns at the excitation frequency are ignored.

In the case of multiple bunches in one ring, the BTF excitation signal takes into account the phase between the bunches to replicate the situation in the real machine where all bunches are excited by the same excitation signal.
\subsubsection{Test of the BTF model}
\query{In Fig.~1, please replace "Amp." by  "Amplitude" and "Phase [$\pi$]" by "Phase$/ \pi$". Also, please check
"$u [1]$. Is simply $u$ intended?}
\begin{figure}[htb]
\includegraphics[width=\columnwidth]{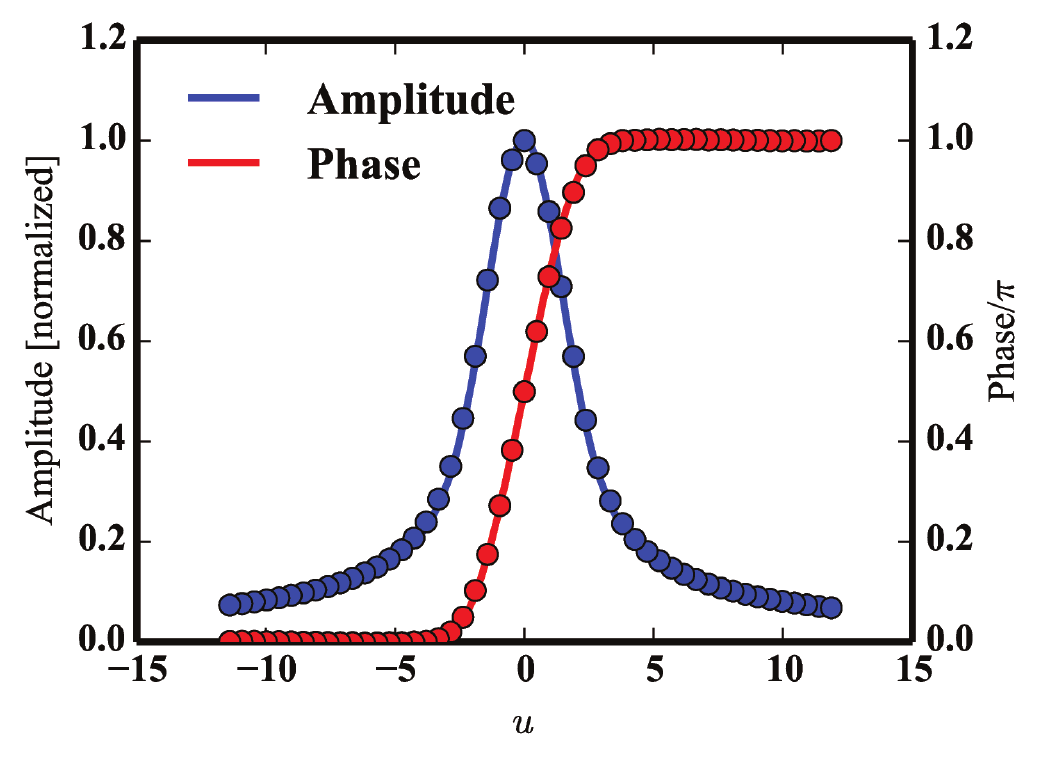}
\caption{Comparison of amplitude and phase of the analytic prediction for the BTF of a beam with Gaussian tune spread (line) and the PIC BTF simulation results (points); $u$ is the normalized frequency coordinate given in eq.~(\ref{eq:normu}). The simulation agrees well with analytic predictions.}\label{fig:anal}
\end{figure}
Before considering BTF of more complex situations we needed to validate our simulation model to make sure we agree with known analytic results. A good benchmark is the BTF of a beam with a Gaussian velocity profile and a tune spread solely due to chromaticity. We consider the single particles as harmonic oscillators around their betatron frequency $\omega$. In the derivation we follow~\cite{ng}. A harmonic oscillator driven off-frequency at a frequency of $\Omega$ carries out a beating at an amplitude $A$ which is proportional to $\frac{1}{\omega^2-\Omega^2}$. To determine the response of an ensemble of harmonic oscillators to a driving frequency $\Omega$ the intuitive approach is to integrate amplitude over the density $\psi(\omega)$ of eigenfrequencies. To make things simpler, $A$ can be approximated by $\frac{1}{2\omega_\beta(\omega-\Omega)}$ where $\omega\simeq\Omega\simeq\omega_\beta$ (which is the case around the betatron lines). Taking out constant factors, the BTF behaves as
\begin{equation}
R(\Omega)\propto\int\frac{1}{\omega-\Omega}\psi(\omega)\,\mathrm d \omega.\label{eq:chrombtf}
\end{equation}
This equation has an analytic solution for different forms of frequency distributions~\cite{ng}. We look at the result for a Gaussian frequency spread. The normalized frequency $u$ can be defined as a function of the mean particle betatron frequency $\overline{\omega}$, the driving frequency $\Omega$ and the frequency width of the distribution $\Delta\omega$ via the equation,
\begin{equation}
u=\frac{\overline \omega- \Omega}{\Delta \omega}.\label{eq:normu}
\end{equation}
Then the real and imaginary parts of the BTF are proportional to $f$ and $g$, as below~\cite{ng}:
\begin{eqnarray}
f(u)&=&\sqrt{\frac{2}{\pi}}e^{-u^2/2}\int_0^\infty\frac{\mathrm dy}{y}e^{-y^2/2}\sinh (u y),\\
g(u)&=&\sqrt{\frac{\pi}{2}}e^{-u^2/2}.
\end{eqnarray}
The analytic BTF compares well to our PIC code as shown on example data in Fig.~\ref{fig:anal}. It is also noteworthy that eq.~(\ref{eq:fullanaBTF}) simplifies to eq.~(\ref{eq:chrombtf}) in the absence of the lens.

\subsection{Tune Distribution}
The tune distributions in simulation are computed by running the simulation without BTF excitation for (2000 to 8000) turns. The particle coordinates are stored for a subset of typically ($10^4$ to $10^5$) particles. In post-processing, DFT is used to find the peak of the oscillation frequency. The analytic tune distributions are computed by numerically evaluating particle density over the distribution and binning the resulting fractions into bins depending on the corresponding tune change according to eq.~(\ref{eq:fullDQ}).

\subsection{Electron-lens Model}
To model the electron lens, the code uses the analytic expressions for the fields of a round Gaussian beam to kick the particles at one of the interaction points. The intensity of the field is adjusted to correspond to a chosen beam--beam parameter. The electron lens in simulation can be run as a \emph{positron lens} by simply changing the sign of the beam--beam parameter, which we did most of the time to be able to easily make a comparison between BTF with stationary Gaussian lens and BTF with split tunes.

\section{Simulation study}
\begin{figure}
\includegraphics[width=\columnwidth]{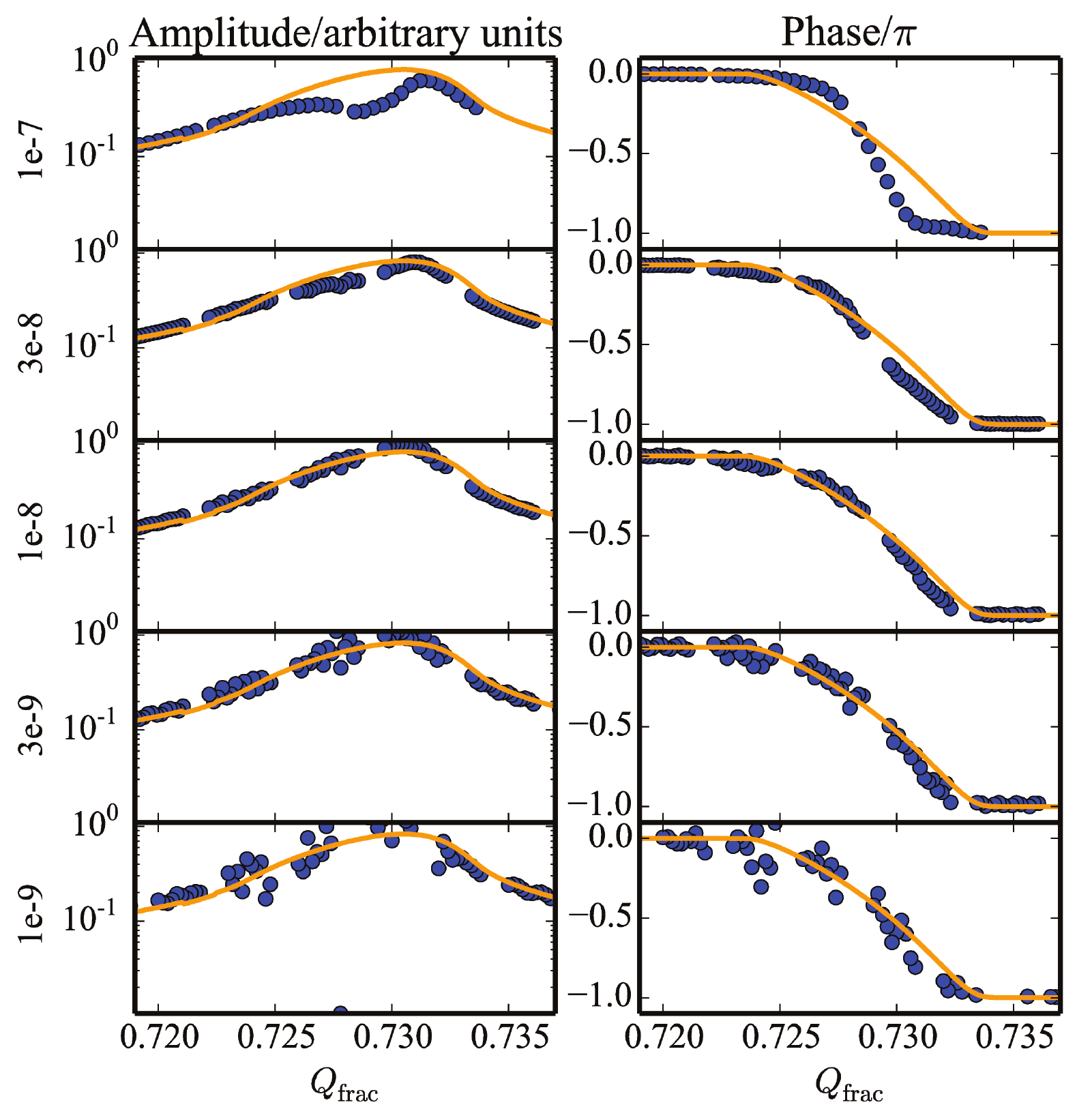}
\caption{Comparison of numeric BTF (dots) from PIC simulation with split tunes and one IP to the analytic expectation (line). Normalized amplitudes are on the left, and values of phase$/\pi$ are on the right. The excitation amplitude (the amplitude of the sinusoidal signal added to the $x'$ component of the particle vector) is given on the left of the plots. We see good agreement for medium amplitudes. For higher amplitudes we observe a deviation, possibly due to particles in the tails of the distribution or due to coherent modes. For lower amplitudes the numerical noise is higher than the signal but can be reduced by increasing particle number.}\label{fig:AmpComp}
\end{figure}

First we ran simulations of a beam undergoing an interaction solely with a Gaussian lens. At the beginning it appeared as though the BTF from simulation and the analytic expectation were in disagreement: the first simulations for both split tunes and electron-lens type configuration showed a double peak structure not present in the analytic expectation. However, as shown in a sweep of excitation amplitudes in Fig.~\ref{fig:AmpComp} on the example of a split tunes simulation, a significant reduction in exciter amplitude led to results in which no double peaks were observed any more. After amplitude reduction, the simulation results for a Gaussian lens were in good agreement with the analytic formula from eq.~(\ref{eq:fullanaBTF}).
\query{In Figs.2--4 and 6, please replace "Amplitude [a.u.]" by, for example,
"Amplitude/atomic units", and "Phase [$\pi$]" by "Phase$/\pi$.}

\subsection{Recovery of the Beam--Beam Parameter}
\begin{figure}
\includegraphics[width=\columnwidth]{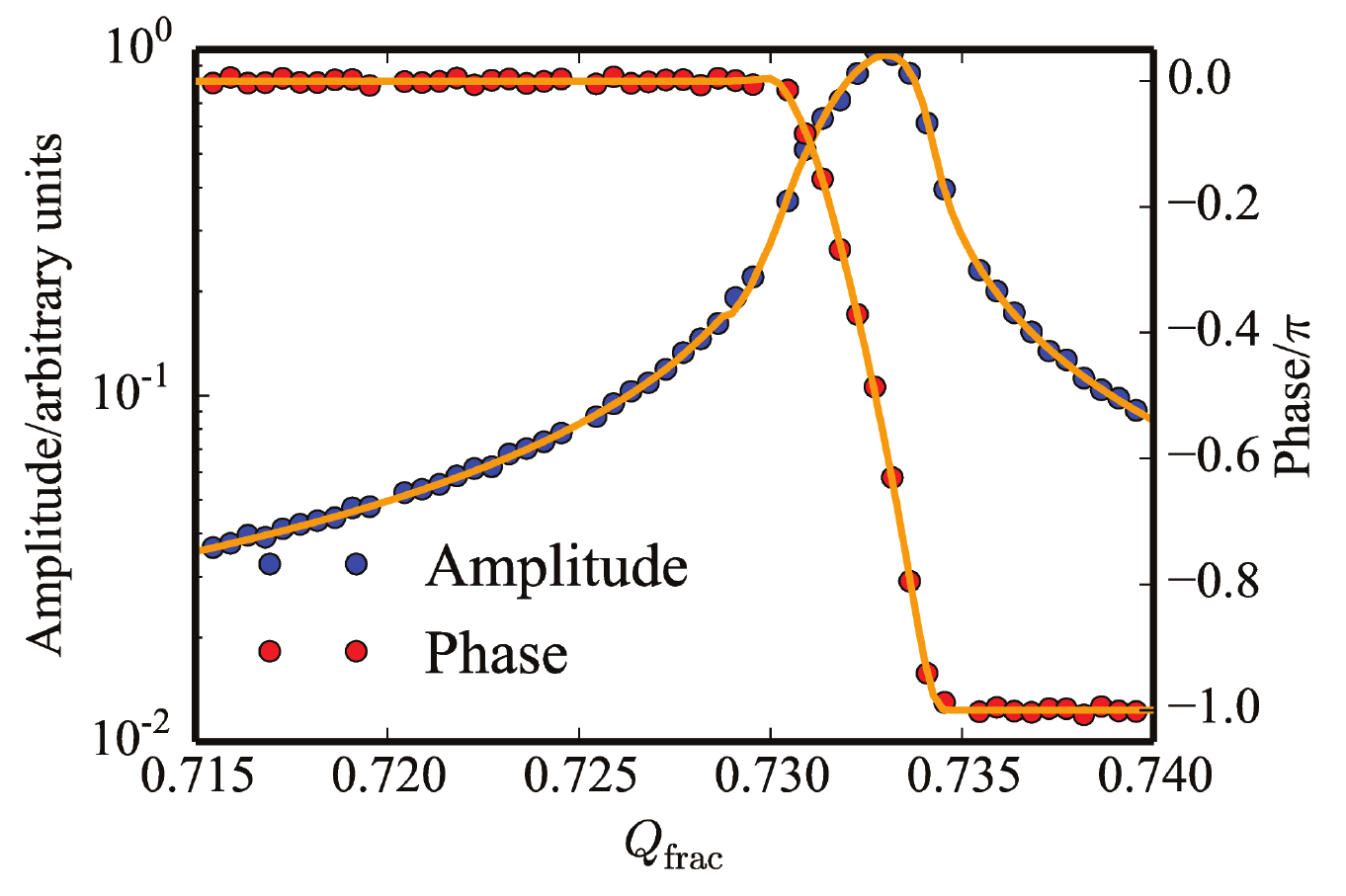}
\caption{Comparison of analytic expectation (orange) with simulation amplitude (blue dots) and phase (red dots) for a Gaussian lens. We see very good agreement.}\label{fig:elensFitExample}
\end{figure}

To test whether fitting to measured BTF of a beam interacting with an electron lens would enable us to recover the beam--beam parameter, we ran simulations of beams with Gaussian lenses of different beam--beam parameters. We fitted the analytic formula for the BTF to simulated BTF. Because the evaluation of the analytic BTF is rather costly, we calculate analytic BTF in \textsc{Mathematica}  for a reasonable range of parameters and use an interpolating function to fit the simulation data. An example fit is shown in Fig.~\ref{fig:elensFitExample}. The beam--beam parameters to which the fits converge lie within three percent of the actual beam--beam parameter chosen for the simulation.  Our test set consisted of simulated BTF with beam--beam parameters between 0.0025 and 0.0145.

\subsection{Split Tune Conditions}
We conducted part of the simulation study on split tune conditions because during the current run the hopes for a running electron lens are not high. The BTF of beams under split tune conditions looked similar to the analytic expectation for a defocussing Gaussian lens.

\begin{figure}
\includegraphics[width=\columnwidth]{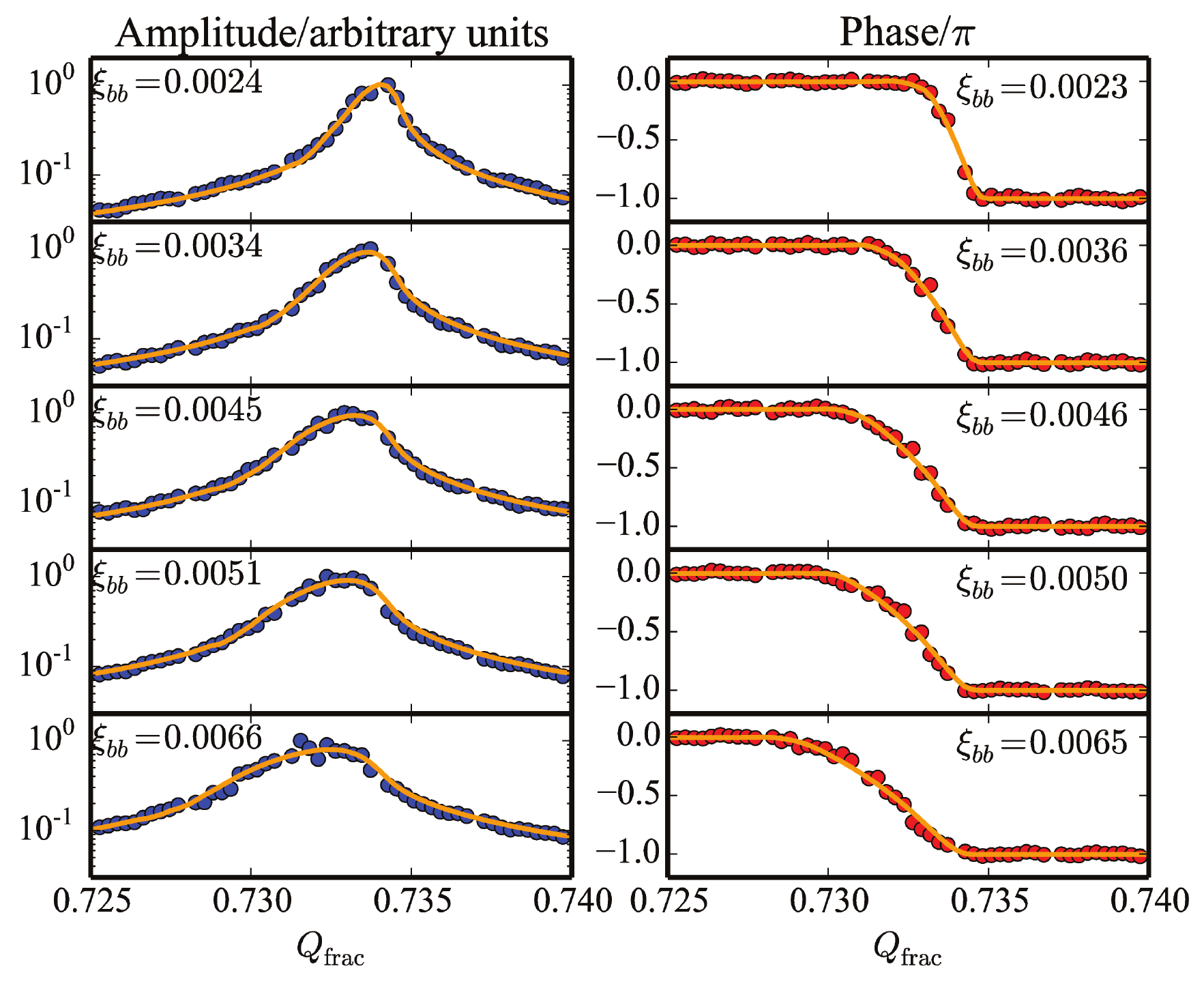}
\caption{BTF from PIC simulations with 1 IP and split tunes (dots) in amplitude (left) and phase (right) for different beam--beam parameters in comparison with fits of the analytic BTF. The result of the fit is given in the individual plots. We see good agreement between the fits to phase and amplitude. Note that the fit results seem to scale linearly with the beam--beam parameter chosen in the simulation but are slightly lower. The dependence of fit result on simulation input is shown in Fig.~\ref{fig:linbbfit}.}\label{fig:fittedExamples}
\end{figure}

\begin{figure}
\includegraphics[width=\columnwidth]{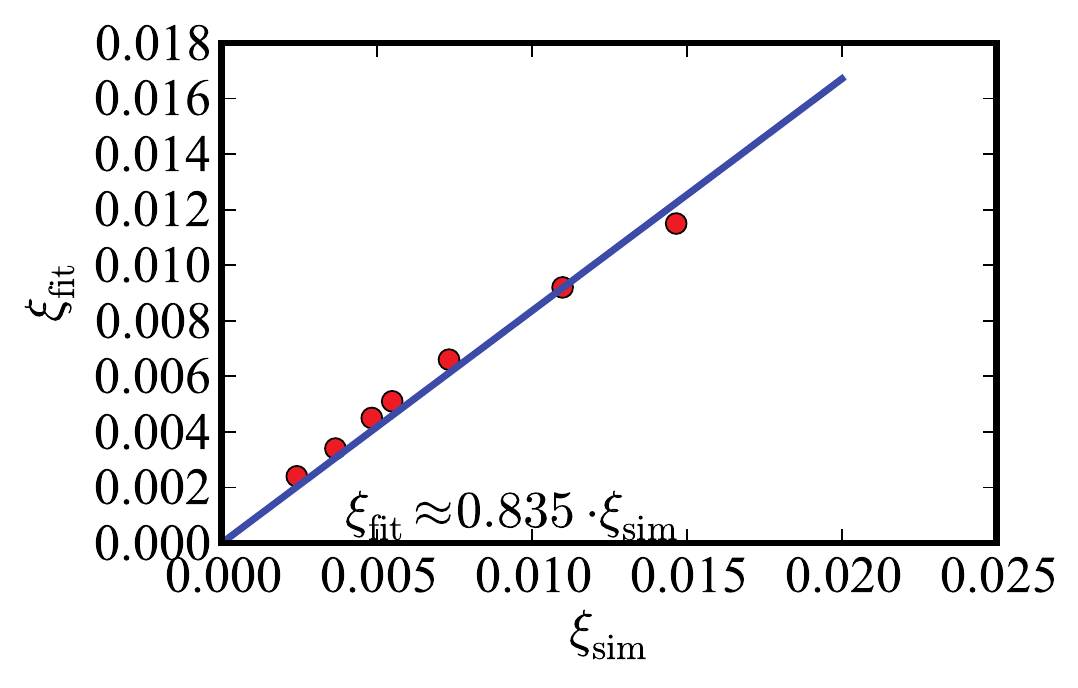}
\caption{Relation between $\xi_\mathrm{bb}$ from fit to BTF simulations of split tunes and from simulation input. The dependence is approximately linear for our range of parameters.}\label{fig:linbbfit}
\end{figure}

We ran simulations for split tune conditions for the same range of beam--beam parameters as we did previously for the electron lens. Again we tried to recover the beam--beam parameters using our fitting routine. For the split tunes we observed a slight deviation of the analytic result from the BTF even though overall agreement was visually still acceptable, as shown in Fig.~\ref{fig:fittedExamples}. The beam--beam parameters recovered from the fits are given in the figure and were slightly underestimating the actual beam--beam parameter used in the simulation. We blame this on the coherent modes to be expected within the incoherent spectrum and possibly leading to a narrowing of the peak. A plot of the beam--beam parameter from the fit over the actual beam--beam parameter from the simulation can be found in Fig.~\ref{fig:linbbfit}. The relation between fit result and actual beam--beam parameter appears to be scaling linearly with a factor of about 0.8. This result might, however, be dependent on the tune splitting.

\subsection{Comparison with Measurement}
\begin{figure}
\includegraphics[width=\columnwidth]{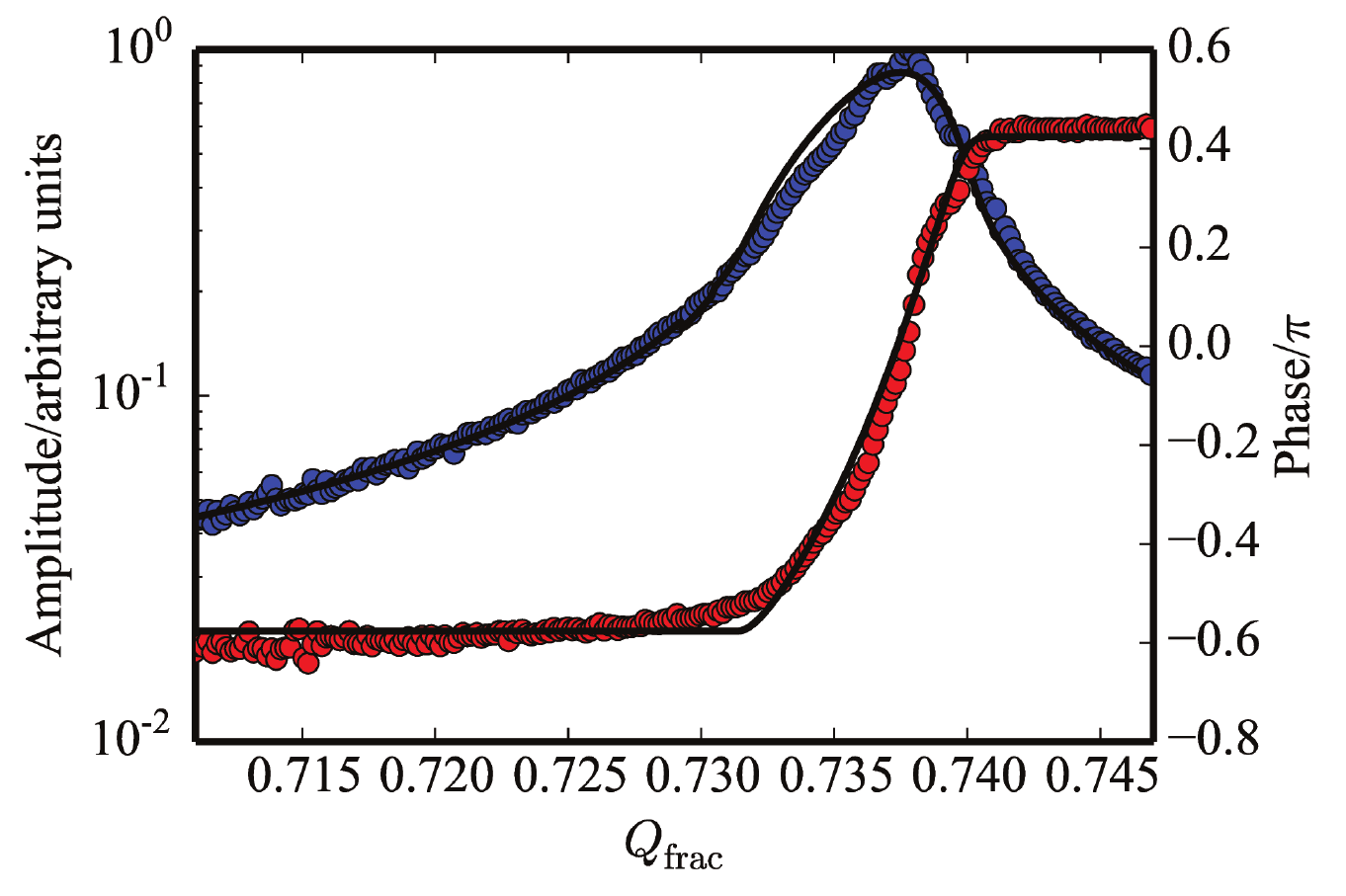}
\caption{Comparison of the fits of the analytic result to measurement data in the yellow horizontal plane. Dots in red are measurement data and those in blue are phase data; the black line is the analytic fit.}\label{fig:measAna}
\end{figure}

The BTF that are available of split tunes measured in the machine do unfortunately not all replicate the analytic expectation. There may be several reasons for that. Firstly the split tunes fills were  full machines (instead of the simulation situation of only two or six coupling bunches). During the 2012 run the BTF system was measuring the
centre-of-mass motion of all the bunches in the machine and combining them into a single BTF. For conditions with differing bunch parameters this may lead to a deformation of the signal so that we cannot expect it to follow our clean simulation data anymore. Furthermore the coherent modes in the incoherent tune distribution can lead to a deformation of the bunches and a resulting deformation of the BTF not covered by our assumption of round Gaussian beams. For this reason we decided to look at the BTF of the best-behaved fills among the split tunes.
Best-behaved means, in this case, no multipeak-structures in the individual planes, low heating compared to the other fills, and rather round beams. The beams were, however, still slightly asymmetrical (normalized 6 $\sigma$ emittances were, for yellow, $\varepsilon_x=20,\quad\varepsilon_y=17.5$, and, for blue, $\varepsilon_x=22.5,\quad\varepsilon_y=21.5$) at
an average of $1.8\cdot10^{11}$ particles per bunch. Nevertheless we tried to apply the fitting algorithm for round beams demonstrated above on simulated BTF. In the horizontal plane according to the beam properties we would expect a beam--beam parameter of 0.012 to 0.014. We found a reasonable  approximation of the measurement by the analytic result. The beam--beam parameters obtained from the fits to yellow and blue horizontal BTF measurement data are $\xi_\mathrm{fit,yellow}=0.012$ and $\xi_\mathrm{fit,blue}=0.012$ when one takes into account the factor of 0.835 between $\xi_\mathrm{fit}$ and $\xi$ obtained from simulation. Furthermore the fits to amplitude and phase deviate only by a few percent. In the vertical plane, the peaks looked distorted, on which basis we reason that here other effects might be at work. On top of that, in the vertical plane, coherent modes were observed for some fills. Example fits in the horizontal plane can be found in Fig.~\ref{fig:measAna}.

\section{Conclusion and outlook}
Currently our analytic model is restricted to round Gaussian beams, as is typically the case in RHIC. However, it should be feasible to generalize the analytic theory for arbitrary aspect ratios by adjusting $\Delta Q_\mathrm{lens}$. In absence of an electron lens a possible test scenario for the fitting method could be found in weak--strong beam--beam interactions, where a strong beam could be modelled as the electron lens and the measurement would be done on the weak beam. For the 2013 run, the BTF system has been upgraded and is now able to measure BTF of single bunches, which would enable running different intensity strong--weak BTF in one fill. Furthermore, once the electron lenses are up and running, we can test whether the BTF of an electron lens agrees with the BTF according to eq.~(\ref{eq:fullanaBTF}), as is to be expected according to our simulation. In this case we would be able to give a good estimate for the strength of the electron lens from the BTF alone.

\section{ACKNOWLEDGMENTS}
Paul G\"orgen would like to thank Simon White for his support.

\end{document}